\documentclass[11pt]{article}
\usepackage[dvips]{graphicx}
\usepackage{mathrsfs}
\usepackage{eucal}
\usepackage{amsmath,amsthm,amssymb}
\usepackage{wrapfig}
\usepackage[dvips]{color}
\usepackage{cite}
\usepackage{framed}
\usepackage{tabularx}
\usepackage[sectionbib]{chapterbib}
\usepackage{threeparttable}
\usepackage{float}
\definecolor{shadecolor}{gray}{0.80}
\DeclareMathVersion{chem}
\SetSymbolFont{letters}{chem}{OT1}{cmr}{m}{n}
\definecolor{shadecolor}{rgb}{0.9, 0.9, 0.90}

\newcommand{\HG}{\hspace{1.3mm}\Hat{\textit{\textsf{\hspace{-1.3mm}G}}}\hspace{0.5mm}}
\newcommand{\I}{I}
\newcommand{\II}{I\hspace{-0.5mm}I}

\begin{document}
\renewcommand{\figurename}{\small{Fig.}~}
\renewcommand{\thefootnote}{$\dagger$\arabic{footnote}}

\begin{flushright}
\textit{Theory of Excluded Volume Effect}
\end{flushright}

\vspace{1mm}
\begin{center}
\setlength{\baselineskip}{25pt}{\LARGE\textbf{Diffusion and Chemical Potential in Polymer Solutions}}
\end{center}

\vspace*{10mm}
\begin{center}
\large{Kazumi Suematsu\footnote{The author takes full responsibility for this article.}, Haruo Ogura\footnote{Kitasato University}, Seiichi Inayama\footnote{Keio University} and Toshihiko Okamoto\footnote{Tokyo University}} \vspace*{2mm}\\
\normalsize{\setlength{\baselineskip}{12pt}
Institute of Mathematical Science\\
Ohkadai 2-31-9, Yokkaichi, Mie 512-1216, JAPAN\\
E-Mail: ksuematsu@icloud.com,  Tel/Fax: +81 (0) 59 326 8052}\\[8mm]
\end{center}

\vspace{3mm}
\hrule
\vspace{4mm}
\noindent\textbf{\large Abstract}\\[2mm]
We give a mathematical proof for the preceding derivation of the excluded volume theory on the basis of the concept of diffusion, chemical potential, and the theory of total differential.

\vspace{3mm}
\noindent\textbf{Key Words}:
\normalsize{Diffusion/ Chemical Potential/ Total Differential/ Excluded Volume Effects/ }\\[0mm]

\hrule
\vspace{3mm}
\setlength{\baselineskip}{13pt}
\section{Introduction}
Generally, diffusion occurs from a concentrated region to a dilute region so as to reduce the concentration gradient. As has been well understood, however, this is not always true. The excluded volume problem of polymers is a good example in which this empirical rule is not applicable. Since monomers are joined by chemical bonds, there is retraction force due to the rubber elasticity, so that the system must maintain the balance between the forward force due to diffusion and the backward force due to elastic retraction. It thus makes possible the reverse flow of segments from the dilute region to the concentrated region against the concentration gradient. For this reason, the complete homogeneity of segment concentration can not always be expected for polymer solutions. In this short paper, we will investigate this problem on the basis of the concept of diffusion and chemical potential, and then give a mathematical proof for the equation derived in the preceding paper\cite{Kazumi}.
\section{Theoretical}
\subsection{Ordinary Diffusion}
\begin{figure}[h]
\begin{center}
\includegraphics[width=11cm]{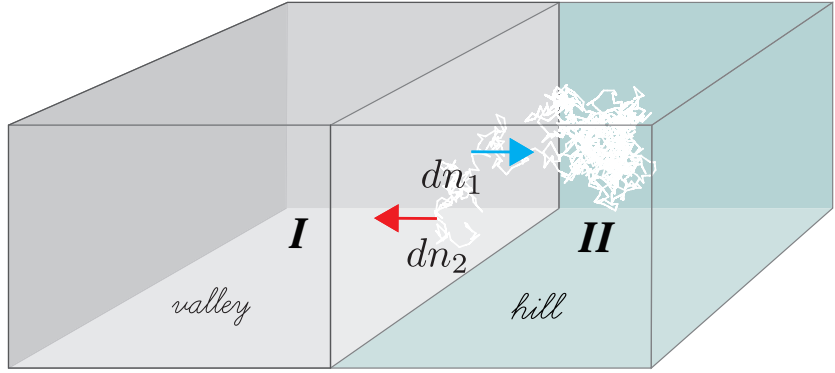}
\caption{Segment migration between the phase I and the  phase II.}\label{SegmentFlow}
\end{center}
\end{figure}

Let $n_{i}$ be the number of molecules of species $i$. The basic thermodynamic equation is
\begin{equation}
d G=V dP- S dT+\sum_{i}\left(\frac{\partial G}{\partial n_{i}}\right)_{T, P, n_{j}}dn_{i}\label{AlD-1}
\end{equation}
Consider a two-component system consisting of $n_{1}$ solvent molecules and $n_{2}$ solute molecules. Suppose that there are two phases in the system, phase I and phase II which have an equal volume size ($V_{\I}=V_{\II}$); for instance, the phase I is rich in $n_{1}$ and II is rich in $n_{2}$. Let $n_{1}$ and $n_{2}$ be able to diffuse freely between these phases. In this paper, we use the word, phase, in a broader meaning; e.g., it does not necessarily mean the abrupt change between two phases such as the $Gas$-$Liquid$ transition. Under constant $T$ and $P$, we have
\begin{equation}
d G=d G_{1}+d G_{2}=\mu_{n_{1\I}} \,d n_{1\I}+\mu_{n_{1\II}} \,d n_{1\II}+\mu_{n_{2\I}} \,d n_{2\I}+\mu_{n_{2\II}} \,d n_{2\II}\label{AlD-2}
\end{equation}
The constraint conditions are $n_{1\I}+n_{1\II}=n_{1}=const.$ and $n_{2\I}+n_{2\II}=n_{2}=const.$, so that 
\begin{align}
d n_{1\I}=&-d n_{1\II}\notag\\
d n_{2\I}=&-d n_{2\II}\label{AlD-3}
\end{align}
Substituting Eq. (\ref{AlD-3}) into Eq. (\ref{AlD-2}), we have
\begin{equation}
d G=\left(\mu_{n_{1\II}}-\mu_{n_{1\I}}\right) \,d n_{1\II}+\left(\mu_{n_{2\II}}-\mu_{n_{2\I}}\right) \,d n_{2\II}\label{AlD-4}
\end{equation}
If the diffusion occurs spontaneously, we must have
\begin{equation}
d G=\left(\mu_{n_{1\II}}-\mu_{n_{1\I}}\right) \,d n_{1\II}+\left(\mu_{n_{2\II}}-\mu_{n_{2\I}}\right) \,d n_{2\II}\le 0\label{AlD-5}
\end{equation}
At equilibrium, $d G=0$. If $d n_{1\II}$ and $d n_{2\II}$ are independent of each other (such as in the case of a mixing process of two ideal gases partitioned by a certain membrane), the equal sign holds for arbitrary $d n_{1\II}$ and $d n_{2\II}$, so that each term in Eq. (\ref{AlD-5}) must be equal to zero. Note that the differential operators, $d n_{1\II}$ and $d n_{2\II}$, are quantities that approach to zero without limit, but not numbers in the ordinary sense, so these cannot be equated with zero. The solution is thus
\begin{align}
\mu_{n_{1\I}}=\mu_{n_{1\II}}\notag\\
\mu_{n_{2\I}}=\mu_{n_{2\II}}\label{AlD-6}
\end{align}
This is the familiar conclusion for the phase equilibria. However, the differential operators $d n_{1\I}$ and $d n_{2\I}$ are, in general, not independent. 

\subsection{Substitutional Diffusion}
Let us consider the simplest case in which a solvent and a solute have the same molecular volume, and that a substitutional diffusion process occurs so that the flow of one solvent molecule from the phase I to the phase II is concurrently accompanied by the displacement of one solute molecule from the phase II to the phase I $-$ the substitutional diffusion, of course, does not exclude the displacement between the same molecular species: solute-solute or solvent-solvent in which the Gibbs potentials of respective phases do not change. The excluded volume problem of a polymer will correspond to this case. In such a case, additional restraint conditions should occur: 
\begin{align}
n_{1\I}+n_{2\I}=const.\notag\\
n_{1\II}+n_{2\II}=const.\label{AlD-7}
\end{align}
so that
\begin{align}
d n_{1\I}=&-d n_{2\I}\notag\\
d n_{1\II}=&-d n_{2\II}\label{AlD-8}
\end{align}
Then, $G$ can be expressed as a one-component-function of $n_{1}$ or $n_{2}$, and we have
\begin{align}
\mu_{n_{2}\I}=&\left(\frac{\partial G}{\partial n_{2\I}}\right)_{T, P}=\left(\frac{\partial G}{\partial n_{1\I}}\right)_{T, P}\frac{dn_{1\I}}{dn_{2\I}}=-\left(\frac{\partial G}{\partial n_{1\I}}\right)_{T, P}=-\mu_{n_{1}\I}\notag\\
\mu_{n_{2}\II}=&\left(\frac{\partial G}{\partial n_{2\II}}\right)_{T, P}=\left(\frac{\partial G}{\partial n_{1\II}}\right)_{T, P}\frac{dn_{1\II}}{dn_{2\II}}=-\left(\frac{\partial G}{\partial n_{1\II}}\right)_{T, P}=-\mu_{n_{1}\II}\label{AlD-9}
\end{align}
which result in the equality:
\begin{equation}
\left(\mu_{n_{1\II}}-\mu_{n_{1\I}}\right) \,d n_{1\II}=\left(\mu_{n_{2\II}}-\mu_{n_{2\I}}\right) \,d n_{2\II}\label{AlD-10}
\end{equation}
If $dG$ is a function of $n_{1}$ or $n_{2}$ alone, then we have $\mu_{n_{1}}=0$ or $\mu_{n_{2}}=0$. This means that either of the terms of the right-hand side in Eq. (\ref{AlD-5}) must be abolished, which yields, for instance,
\begin{equation}
d G=\left(\mu_{n_{2\II}}-\mu_{n_{2\I}}\right) \,d n_{2\II}\label{AlD-11}
\end{equation}
This is a natural consequence, because the two opposite flows of $n_{1}\,(\I\rightarrow \II)$ and $n_{2}\,(\II\rightarrow \I)$ concurrently occur.

\subsection{Excluded Volume Problem}
The above discussion must be modified for the excluded volume problem in polymer solutions, since, in addition to the diffusive migration of segments, there is retraction force due to the rubber elasticity of a polymer. The total free energy of a polymer must be written as $G=G_{\text{diffusion}}+G_{\text{elasticity}}$, and it must satisfy
\begin{equation}
d G=dG_{\text{diffusion}}+dG_{\text{elasticity}}\le 0\label{AlD-12}
\end{equation}
In this problem, what is essential is the force balance between the expansion and the retraction of a polymer; so the chemical potential of solvent need not be taken into account. By Eq. (\ref{AlD-11}), this may be expressed in the form:
\begin{equation}
d G=dG_{\text{diffusion}}+dG_{\text{elasticity}}=\left(\mu_{n_{2\II}}-\mu_{n_{2\I}}\right) \,d n_{2\II}+dG_{\text{elasticity}}\le 0\label{AlD-13}
\end{equation}
At equilibrium, $dG$ must take the minimum value.

\subsubsection{$dG_{\text{diffusion}}$}
The fundamental equation\cite{Flory} of the Gibbs potential of mixing pure solvent and pure polymer melt is:
\begin{equation}
\Delta G_{mixing}=\,\frac{kT}{V_{1}}\int\left\{-\left(1-\chi\right)\Hat{v}_{2}+\left(1/2-\chi\right)\Hat{v}_{2}^{2}+\frac{1}{6}\Hat{v}_{2}^{3}+\cdots\right\}\delta V\label{AlD-14}
\end{equation}
where $V$ denotes the system volume and $V_{1}$ the volume of a solvent molecule, and $\Hat{v}_{2}$ represents the volume fraction of polymer segments in the local area $\delta V$. Let $N$ be the number of segments constituting a polymer. Assuming the Gaussian distribution of segments around the center of gravity, $\Hat{v}_{2}$ can be expressed in the form:
\begin{align}
\Hat{v}_{2}=&V_{2}\Hat{C}=V_{2}N\left(\frac{\beta}{\pi\alpha^{2}}\right)^{3/2}\sum_{\{a, b, c\}}\exp\left\{-\frac{\beta}{\alpha^{2}}\left[(x-a)^{2}+(y-b)^{2}+(z-c)^{2}\right]\right\}\notag\\
\equiv&V_{2}N\left(\frac{\beta}{\pi\alpha^{2}}\right)^{3/2}\HG(x, y, z)\label{AlD-15}
\end{align}
where $V_{2}$ denotes the volume of a segment, and $\beta=3/2\langle s_{N}^{2}\rangle_{0}$ has the usual meaning (the subscript 0 denotes the unperturbed dimensions); $\{a, b, c\}$ signifies the location of individual polymers, so that $\Hat{C}$ represents the number concentration of segments at the coordinate $(x, y, z)$ with the symbol \,$\Hat{}$\, denoting the Gaussian approximation specified by Eq. (\ref{AlD-15}). In this problem, $n_{1}$ corresponds to solvent molecules, and $n_{2}$ corresponds to segments. Note that $\Hat{C}$ is a function of $\Hat{v}_{2}$; so $\Delta G_{mixing}$ is a one-variable-function of $n_{2}$. As we can see from Eq. (\ref{AlD-15}), a polymer solution is an inhomogeneous system of the segment concentration. The inhomogeneity causes physical instability of the solution. To acquire the stability, a polymer must reduce the inhomogeneity by expanding the mean radius and making its segments overlap with those of other molecules.

Because of the wild inhomogeneity of segment concentration, we may regard a polymer solution as consisting of two phases, a concentrated region (phase II) and a dilute region (phase I) which is put in contact with each other.

\vspace{5mm}
\hrule height 0.7pt width 10cm
\vspace{3mm}
To proceed with our discussion, it is useful to modify the basic thermodynamic equation\cite{Kazumi}. Multiply Eq. (\ref{AlD-1}) by $V/V$, and we have
\begin{equation}
d G=V dP- S dT+\sum_{i}\left(\frac{\partial G}{\partial c_{i}}\right)_{T,P}dc_{i}\label{AlD-16}
\end{equation}
where $c_{i}$ represents the number concentration of molecular species $i$. Thus we can introduce a new definition of the chemical potential as a measure of the rate of the change of Gibbs potential as against the change of solute concentration\cite{deGennes} under constant $T$ and $P$:
\begin{equation}
\mu_{c_{i}}=\left(\frac{\partial G}{\partial c_{i}}\right)_{T,P}\label{AlD-17}
\end{equation}
\vspace{3mm}
\hrule height 0.7pt width 10cm

\vspace{5mm}
Using Eq. (\ref{AlD-17}), Eq. (\ref{AlD-13}) may be recast in the form:
\begin{equation}
d G=dG_{\text{diffusion}}+dG_{\text{elasticity}}=\left(\mu_{c_{2\II}}-\mu_{c_{2\I}}\right) d c_{2\II}+dG_{\text{elasticity}}\label{AlD-18}
\end{equation}
The equilibrium point is determined by the force balance between the expansion of a molecule due to the diffusive force of segments and the retraction force due to the rubber elasticity; it corresponds to the minimum point of $G$. To find the equilibrium point, differentiate Eq. (\ref{AlD-18}) with respect to the expansion factor, $\alpha$, to yield
\begin{equation}
\frac{d G}{d\alpha}=\left(\mu_{c_{2\II}}-\mu_{c_{2\I}}\right)\frac{d c_{2\II}}{d\alpha}+\frac{dG_{\text{elasticity}}}{d\alpha}=0\label{AlD-19}
\end{equation}
Eq. (\ref{AlD-19}) is a basic equation for the excluded volume effects of polymer solutions.

To apply Eq. (\ref{AlD-19}) to real problems, what we must do is only to express this equality in terms of polymer solutions, making one-to-one correspondence of the notations between the thermodynamic formula (Table \ref{AlD-Table1}), namely
\begin{equation}
\mu_{c_{2}}=\left(\frac{\partial G}{\partial \Hat{v}_{2}}\right)_{T,P}\frac{d\Hat{v}_{2}}{d\Hat{C}}\label{AlD-20}
\end{equation}
where $\Hat{v}_{2}$ is given by Eq. (\ref{AlD-15}), and 
\begin{equation}
\Hat{C}=N\left(\frac{\beta}{\pi\alpha^{2}}\right)^{3/2}\HG(x, y, z)\label{AlD-21}
\end{equation}
Eq. (\ref{AlD-20}) is common to both the phases I and II, since the specification of the phases is determined only through the integral operation with respect to $dV=dxdydz$. By Eqs. (\ref{AlD-14}) and (\ref{AlD-15}), $d\Hat{v}_{2}/d\Hat{C}=V_{2}$ and
\begin{align}
\mu_{c_{2}}=\left(\frac{\partial G}{\partial \Hat{v}_{2}}\right)_{T,P}\frac{d\Hat{v}_{2}}{d\Hat{C}}=\,kT\frac{V_{2}}{V_{1}}\iiint\left\{-(1-\chi)+\left(1-2\chi\right)\Hat{v}_{2}+\frac{1}{2}\Hat{v}_{2}^{2}+\cdots\right\}dxdydz\label{AlD-22}
\end{align}
$\partial\Hat{C}/\partial\alpha$ can be calculated directly using Eq. (\ref{AlD-21}) to yield
\begin{equation}
\left(\frac{\partial\Hat{C}}{\partial \alpha}\right)=-\frac{3}{\alpha}\frac{\Hat{v}_{2}}{V_{2}}+\frac{2\beta N}{\alpha^{3}}\left(\frac{\beta}{\pi\alpha^{2}}\right)^{3/2}\sum_{\{a, b, c\}}s^{2}(x, y, z, a, b, c)\exp\left\{-\frac{\beta}{\alpha^{2}}s^{2}(x, y, z, a, b, c)\right\}\label{AlD-23}
\end{equation}
where $s^{2}(x, y, z, a, b, c)=(x-a)^{2}+(y-b)^{2}+(z-c)^{2}$.

\begin{center}
  \begin{threeparttable}[h]
    \caption{One-to-one Correspondence of Notations}\label{AlD-Table1}
  \begin{tabular}{c l}
\hline\\[-1.5mm]
Thermodynamic equation & Excluded volume problem \\[2mm]
\hline\\[-1.5mm]
$n_{1}$ & Number of solvent molecules\\[1.5mm]
$n_{2}$ & Number of segments ($N$)\\[1.5mm]
$c_{2}$ & Segment concentration ($\Hat{C}$)\\[1.5mm]
$v_{1}$ & Fraction of solvent ($v_{1}$)\\[1.5mm]
$v_{2}$ & Fraction of segments ($\Hat{v}_{2}$)\\[1.5mm]
Phase $\text{I}$ & \textit{valley}: Outside of a polymer\\[1.5mm]
Phase $\text{II}$ & \textit{hill}: Inside of a polymer\\[1.5mm]\\[-3mm]
\hline\\[-6mm]
   \end{tabular}
    \vspace*{2mm}
  \end{threeparttable}
  \vspace*{4mm}
\end{center}

\subsubsection{$dG_{\text{elasticity}}$}
The expression for the elastic potential is already given in the previous works \cite{Flory, Treloar, Kazumi}:
\begin{equation}
\left(\frac{\partial\Delta G_\text{elasticity}}{\partial \alpha}\right)_{T,P}=-T\left(\frac{\partial \Delta S}{\partial \alpha}\right)_{T, P}=3kT\left(\alpha-1/\alpha\right)\label{AlD-24}
\end{equation}

\subsubsection{Integration of the Osmotic and the Elastic Terms}
Substituting Eqs. (\ref{AlD-22}) and (\ref{AlD-24}) into Eq. (\ref{AlD-19}), and taking difference between the phase I and the phase II, we have
\begin{multline}
\alpha-1/\alpha =-\frac{V_{2}}{3V_{1}}\left\{\iiint\left(\left(1-2\chi\right)\Hat{v}_{2\II}+\frac{1}{2}\Hat{v}_{2\II}^{2}+\cdots\right)\left(\frac{\partial\Hat{C}}{\partial \alpha}\right)_{\hspace{-0.3mm}\II} dx dy dz\right.\\
\left.-\iiint\left(\left(1-2\chi\right)\Hat{v}_{2\I}+\frac{1}{2}\Hat{v}_{2\I}^{2}+\cdots\right)\left(\frac{\partial\Hat{C}}{\partial \alpha}\right)_{\hspace{-0.3mm}\I} dx dy dz\right\}\label{AlD-25}
\end{multline}
In Eq. (\ref{AlD-25}), we have made use of the aforementioned assumption that the phases I and II have an equal volume ($V_{\I}=V_{\II}$). Since the change in order of the integral and the differentiation does not change the result, Eq. (\ref{AlD-25}) may be recast in the form:
\begin{multline}
\alpha-1/\alpha =-\frac{1}{3V_{1}}\frac{\partial}{\partial \alpha}\left\{\iiint\left(\left(\tfrac{1}{2}-\chi\right)\Hat{v}^{2}_{2\II}+\frac{1}{6}\Hat{v}_{2\II}^{3}+\cdots\right) dx dy dz\right.\\
\left.-\iiint\left(\left(\tfrac{1}{2}-\chi\right)\Hat{v}^{2}_{2\I}+\frac{1}{6}\Hat{v}^{3}_{2\I}+\cdots\right) dx dy dz\right\}\label{AlD-26}
\end{multline}
which is exactly the result in the preceding paper\cite{Kazumi}, if we alter the notations as $\I\rightarrow valley$ and $\II\rightarrow hill$. As a consequence, the preceding formula derived intuitively has gained a mathematical basis.


\end{document}